\documentstyle[aps]{revtex}
\begin{document}
\draft
\title{
Parametric amplification with a friction in heavy ion collisions
}
\author{Masamichi Ishihara
\thanks{Electric address: m\_isihar@nucl.phys.tohoku.ac.jp}
}
\address{Department of Physics, Tohoku University \\
Aoba-ku, Sendai 980-8578, Japan 
}
\date{\today}
\maketitle
\begin{abstract}
We study the effects of the expansion of the system  and the friction
on the parametric amplification of mesonic fields 
in high energy heavy ion collisions within the linear $\sigma$ model .
The equation of motion which is similar to Mathieu equation 
is derived to describe the time development of the fields 
in the last stage of a heavy ion collision after the freezeout time.
The enhanced mode is extracted analytically by 
comparison with Mathieu equation
and 
the equation of motion is solved numerically to examine
whether soft modes will be enhanced or not.
It is found that
a strong peak appears around 267 MeV in the pion transverse momentum distribution 
in cases with weak friction and high maximum temperature.
This enhancement may be extracted by taking the ratio between different modes
in the pion transverse momentum distribution.
\end{abstract}


\section{Introduction}
The chiral symmetry restoration has been investigated theoretically
and is expected to be realized in the coming experiments 
of high energy heavy ion collisions  at RHIC and LHC.
The chiral symmetry restoration leads to the subsequent rebreaking 
which has been studied in terms of the interesting phenomena called 
'Disoriented Chiral Condensates'(DCC). 
DCC is a (quasi) ground state and 
is characterized by finite pion condensates in the linear $\sigma$ model.
Though DCC consists of zero mode in an ideal case,
it consists of soft modes 
in a realistic case.
In the conventional formation scenarios of DCC, it is assumed that 
the condensate rolls down to any direction with equal probability 
from the top of the hill of the mexican hat potential
if the thermal equilibrium above the critical temperature is completely established.
However, this is not always valid in heavy ion collisions at high energies
because the initial condition is such that the condensate is settled on the minimum of the 
potential at zero temperature before the collision.  
We have shown that the condensate moves almost along the sigma axis 
when the effects of random forces are negligible 
\cite{Ishihara2,Ishihara3}.  
In such cases, 
the condensate moves to $\sigma$ (normal) or $-\sigma$ direction 
which is the opposite side to the normal vacuum. 
DCC does not appear if the condensate stays at the normal vacuum, 
while if it stays at $-\sigma$, then it is just a DCC.

As mentioned above, 
the condensate may oscillate along the sigma axis 
around the minimum of the effective potential at later time in high energy heavy ion collisions
because the condensate moves along the sigma axis.
The field which couples to the oscillating field may be amplified due to the so called 
'parametric amplification' \cite{Muller,Hiro-Oka,Kaiser,Dumitru}.
It was pointed out that the enhancement of nonzero mode due to parametric amplification 
on the normal vacuum may occur when the condensate oscillates around 
the minimum of the potential at zero temperature \cite{Muller}.
It was also shown that a squeezed state is formed by parametric amplification 
\cite{Hiro-Oka}.
However,
it is not well-known whether the enhancement of nonzero mode occurs or not
when the amplitude of the oscillation decreases because of the expansion and the friction. 

The equation of motion for the condensate 
with a friction term was derived by Bir\'o and Greiner 
\cite{Biro}  on the chirally symmetric vacuum
where $T > T_{c}$ with $T$ being the temperature of the system 
and $T_{c}$ the critical temperature, 
and by Rischke \cite{Rischke} on chirally broken vacuum where $T < T_{c}$. 
The magnitude of the friction for the condensate 
on the chirally broken vacuum in the linear $\sigma$ model was given by Rischke.
There is a finite friction for $\sigma$ field 
because of $\sigma \rightarrow 2\pi$ decay.
Then, it is better to include the friction in the calculation of 
parametric amplification for soft modes like DCC formation and decay studies 
\cite{Biro,Rischke,withfric,DecayRate}.

The aim of this paper is to investigate 
whether the enhancement of nonzero modes occurs or not 
in one dimensional scaling case when there is a friction.
This paper is organized as follows. 
In sec.\ref{sec:Deriv}, 
the equation of motion is derived and 
a constant friction is introduced to describe the dissipative nature effectively.
After that, the equation of motion is analyzed by 
comparing with Mathieu equation.
In sec.\ref{sec:numerical},
The initial condition is given and 
a measure to extract the amplification of the fields is introduced.
The equation of motion is solved numerically to see 
whether the enhancement of soft modes occurs or not. 
Sec.\ref{sec:conclude} is assigned for conclusions and discussions.

\section{Equation of motion in one dimensional scaling case}
\label{sec:Deriv}
We use the linear $\sigma$ model to describe the parametric amplification
in the last stage (after freezeout) of high energy heavy ion collisions. 
The 
Lagrangian is 
\begin{equation}
{\cal L} = \frac{1}{2} \partial_{\mu} \phi \partial^{\mu} \phi - V(\phi)
, \hspace{2cm}
V(\phi) = \frac{\lambda}{4} \left( \phi^{2}  - v^{2} \right)^{2} - H \sigma',  
\end{equation}
where $\phi = (\sigma',\vec{\pi})$.
The minimum of the potential is determined by differentiating $V(\phi)$: 
\begin{equation}
\left. \frac{\partial V(\phi)}{\partial \sigma'} 
\right|_{\sigma'=f_{\pi},\vec{\pi}=0} = 0 ,
\end{equation} 
where $f_{\pi}$ is the pion decay constant. 
Since we are interested in the motion around the minimum of the potential,
the new field $\sigma = \sigma' - f_{\pi}$ is introduced and 
the potential  is rewritten.
We have the relations: 
\begin{eqnarray} 
m_{\sigma}^{2} \stackrel{\rm def}{=} \lambda (3 f_{\pi}^{2} - v^{2}) 
, \hspace{1cm}
m_{\pi}^{2} \stackrel{\rm def}{=} \lambda (f_{\pi}^{2} - v^{2}) 
, \hspace{1cm}
H = \lambda f_{\pi} (f_{\pi}^{2} - v^{2}),
\end{eqnarray} 
where $m_{\sigma}$ is the mass of the sigma meson and  $m_{\pi}$ is that of the pion. 
The equation of motion for $\sigma$ and $\pi_{i} (i=1,2,3)$ become 
{
\setcounter{enumi}{\value{equation}}
\addtocounter{enumi}{1}
\setcounter{equation}{0}
\renewcommand{\theequation}{\theenumi\alph{equation}}
\begin{eqnarray}
&& 
\Box \sigma(x) + m_{\sigma}^{2} \sigma(x) 
+ \lambda \left( 
\sigma^{2}(x) + 3 f_{\pi} \sigma(x) + \vec{\pi}^{2}(x) 
\right) \sigma(x) + \lambda f_{\pi} \vec{\pi}^{2} = 0 ,
\\ 
&&
\Box \pi_{i}(x) + m_{\pi}^{2} \pi_{i}(x)
+ \lambda \left(
\sigma^{2}(x) + 2 f_{\pi} \sigma(x) + \vec{\pi}^{2}(x)
\right) \pi_{i}(x) = 0 ,
\end{eqnarray}
where $\Box = \partial_{t}^{2} - \partial_{x}^{2} - \partial_{y}^{2}  - \partial_{z}^{2} $.
We solve the above equations of motion, especially for soft modes in the presence of 
the background oscillation of the condensation (zero mode) .  
Since the sum of hard modes generates a friction \cite{Biro,Rischke,Greiner} in general, 
a friction term is introduced phenomenologically.
The equations of motion become
\setcounter{equation}{\value{enumi}}
}
{
\setcounter{enumi}{\value{equation}}
\addtocounter{enumi}{1}
\setcounter{equation}{0}
\renewcommand{\theequation}{\theenumi\alph{equation}}
\begin{eqnarray}
&& 
\Box \sigma(x) + \eta_{\sigma}(x) \partial_{\tau} \sigma(x) + m_{\sigma}^{2} \sigma(x) 
+ \lambda \left(
\sigma^{2}(x) + 3 f_{\pi} \sigma(x) + \vec{\pi}^{2}(x)
\right) \sigma(x) + \lambda f_{\pi} \vec{\pi}^{2} = 0 ,
\label{eqn:deriv_sigma} \\ 
&&
\Box \pi_{i}(x) + \eta_{\pi}(x) \partial_{\tau} \pi(x) + m_{\pi}^{2} \pi_{i}(x)
+ \lambda \left(
\sigma^{2}(x) + 2 f_{\pi} \sigma(x) + \vec{\pi}^{2}(x)
\right) \pi_{i}(x) = 0 ,
\label{eqn:deriv_pi}
\end{eqnarray}
where $\eta_{\sigma}(x)$ and $\eta_{\pi}(x)$ are the frictions for $\sigma$ and $\pi$
fields, respectively.
Convenient variables $\tau$ and $\eta$ are defined by
\setcounter{equation}{\value{enumi}}
}
\begin{eqnarray}
\tau \stackrel{\rm def}{=} \sqrt{t^{2} - z^{2}} ,
\hspace{2cm}
\eta     \stackrel{\rm def}{=} \frac{1}{2} \ln \left( \frac{t+z}{t-z} \right),
\label{def:variables}
\end{eqnarray}
for one dimensional scaling case.
D'Alembertian is rewritten by using these variables:
\begin{equation}
\Box = \partial_{\tau}^{2} + \frac{1}{\tau}\partial_{\tau} 
- \frac{1}{\tau^{2}} \partial_{\eta}^{2} - \partial_{\perp}^{2}, 
\end{equation}
where $\partial_{\perp}^{2} = \partial_{x}^{2} + \partial_{y}^{2} $. 
We assume that the frictions $\eta_{\sigma}$ and $\eta_{\pi}$ depend on only $\tau$.
First, we solve eq.(\ref{eqn:deriv_sigma}) without the interaction term.
The notation $\sigma^{(0)}(\tau)$ which implies zero mode (the condensate) is introduced.
It is the background field for nonzero modes.  
The equation becomes
\begin{eqnarray} 
&& 
\left[ 
\partial_{\tau}^{2} + \left( \eta_{\sigma}(\tau) + \frac{1}{\tau} \right) \partial_{\tau} 
+ m_{\sigma}^{2} 
\right] \sigma^{(0)}(\tau) = 0 , 
\label{eqn:order0_sigma}
\end{eqnarray}
where it is assumed that $\sigma^{(0)}$ is a function of only $\tau$.
Without loss of generality, one can substitute the following factorized form
of $\sigma^{(0)}(\tau)$  into eq.(\ref{eqn:order0_sigma}):
\begin{equation} 
\sigma^{(0)}(\tau)  =  f^{(0)}(\tau) u_{\sigma}^{(0)}(\tau) . 
\label{eqn:div}
\end{equation}
$du_{\sigma}^{(0)}(\tau)/d\tau$ term  appears on the right hand side of 
eq.(\ref{eqn:order0_sigma}) 
and its coefficient vanishes if $f^{(0)}(\tau)$ satisfies the following 
equation: 
\begin{equation}
2 \frac{df^{(0)}(\tau)}{d\tau} + \left( \eta_{\sigma}(\tau) + \frac{1}{\tau} \right) f^{(0)}(\tau) = 0 .
\label{eqn:for_f}
\end{equation}
The equation for $u_{\sigma}^{(0)}(\tau)$ becomes 
\begin{equation}
\frac{d^{2}u_{\sigma}^{(0)}(\tau)}{d\tau^{2}} + \left\{
\frac{1}{2\tau^{2}} - \frac{1}{2}\frac{d\eta_{\sigma}(\tau)}{d\tau} 
- \frac{1}{4}\left(\eta_{\sigma}(\tau)+\frac{1}{\tau}\right)^{2}+ m_{\sigma}^{2}
\right\} u_{\sigma}^{(0)}(\tau) = 0 .
\label{eqn:reduced_sigma0}
\end{equation}
For a large $\tau$,  $1/\tau^{2}$ and $d\eta_{\sigma}/d\tau$ are negligible
if $\eta_{\sigma}$ reaches a constant. 
Then, eq.(\ref{eqn:reduced_sigma0}) for large $\tau$ is approximately
\begin{equation}
\frac{d^{2}u_{\sigma}^{(0)}(\tau)}{d\tau^{2}} 
+ \left\{m_{\sigma}^{2} - \frac{1}{4} \eta_{\sigma}^{2}(\infty)\right\} u_{\sigma}^{(0)}(\tau) = 0. 
\label{eqn:for_v}
\end{equation}
The solutions of eqs.(\ref{eqn:for_f}) and (\ref{eqn:for_v}) are 
{
\setcounter{enumi}{\value{equation}}
\addtocounter{enumi}{1}
\setcounter{equation}{0}
\renewcommand{\theequation}{\theenumi\alph{equation}}
\begin{eqnarray}
f^{(0)}(\tau) &=& C \tau^{-1/2} \exp \left(-\frac{1}{2}\int^{\tau} ds \eta_{\sigma}(s)\right) 
\sim C \tau^{-1/2} \exp \left(- \frac{1}{2} \eta_{\sigma}(\infty) \tau \right) ,
\label{sol:f}
\\
u_{\sigma}^{(0)}(\tau) &=& 
u_{0} \cos \left[
\left( m^{2}_{\sigma}  - \frac{\eta^{2}_{\sigma}(\infty)}{4} \right)^{1/2} \tau + \theta
\right] ,
\label{sol:v}
\end{eqnarray}
where $C$, $u_{0}$ and $\theta$ are constants. 
Then, $O(\lambda^{0})$ solution of $\sigma^{(0)}(\tau) =f^{(0)}(\tau) u_{\sigma}^{(0)}(\tau)$ 
is given with the help of eqs.(\ref{sol:f}) and (\ref{sol:v}).
We can obtain the same expression for $\pi$ fields.
However, we now consider the case in which the motion of $\pi$ field is negligible,
namely $\pi_{i}^{(0)} \sim 0$, because the condensate moves almost along the sigma axis.
\setcounter{equation}{\value{enumi}}
}

We solve eqs.(\ref{eqn:deriv_sigma}) and (\ref{eqn:deriv_pi})  
with the interaction term in order to evaluate 
the effect of the motion of the condensate. 
Fields are decomposed to substract the background oscillation as follows: 
{
\setcounter{enumi}{\value{equation}}
\addtocounter{enumi}{1}
\setcounter{equation}{0}
\renewcommand{\theequation}{\theenumi\alph{equation}}
\begin{eqnarray}
\sigma(x) &=& \sigma^{(0)}(x) + \sigma^{(1)}(x),
\label{eqn:si_O_lambda}
\\
\sigma^{(0)}(x) &=& 
\sigma_{0} \tau^{-1/2} e^{-\frac{1}{2} \eta_{\sigma}(\infty)\tau} 
\cos \left[
\left( m^{2}_{\sigma}  - \frac{\eta^{2}_{\sigma}(\infty)}{4} \right)^{1/2} \tau + \theta
\right],
\label{eqn:init_sigma0}
\\
\pi(x) &=& \pi^{(0)}(x) + \pi^{(1)}(x) \sim \pi^{(1)}(x) , 
\label{eqn:pi_O_lambda}
\end{eqnarray}
where $\sigma_{0} = C u_{0}$ which is a constant.
With the help of eqs.(\ref{eqn:si_O_lambda}) and (\ref{eqn:pi_O_lambda}), we have
\setcounter{equation}{\value{enumi}}
}
{
\setcounter{enumi}{\value{equation}}
\addtocounter{enumi}{1}
\setcounter{equation}{0}
\renewcommand{\theequation}{\theenumi\alph{equation}}
\begin{eqnarray}
\sigma^{2} &=& \left( \sigma^{(0)} \right)^{2} + 2 \sigma^{(0)} \sigma^{(1)} 
+  \left( \sigma^{(1)} \right)^{2} 
= O \left(\left[ \sigma^{(0)} \right]^{2} \right) + 2 \sigma^{(0)} \sigma^{(1)} + O(\lambda^{2}) ,
\label{eqn:expan1}
\\ 
\pi^{2} &=& O(\lambda^{2}) 
\label{eqn:expan15}
\\
\sigma^{3} &=& \left( \sigma^{(0)} \right)^{3} 
+ 3 \left( \sigma^{(0)} \right)^{2} \sigma^{(1)} 
+ 3 \sigma^{(0)} \left( \sigma^{(1)} \right)^{2} 
+  \left( \sigma^{(1)} \right)^{3}
\nonumber \\ 
&=& 
O \left( \left[ \sigma^{(0)} \right]^{3} \right) + O \left( \left[ \sigma^{(0)} \right]^{2} \lambda \right) 
  + O \left( \sigma^{(0)} \lambda^{2} \right) + O(\lambda^{3}) , 
\label{eqn:expan2}
\\
\vec{\pi}^{2} \sigma &=&  O(\lambda^{2}) \times \left( \sigma^{(0)} + O(\lambda) \right) .
\label{eqn:expan3}
\end{eqnarray}
\setcounter{equation}{\value{enumi}}
}
$O \left( \left[ \sigma^{(0)} \right]^{2} \right)$ and higher contributions 
are negligible compared with $O(\sigma^{(0)})$ 
because $\sigma_{0}$ is taken to be small and 
$\tau^{-1/2} \exp \left(-\eta_{\sigma}(\infty) \tau /2 \right)$ is small for large $\tau$.
The lowest contribution of the interaction terms is obtained by  
neglecting $O(\lambda^{2})$ and higher terms in 
eq.(\ref{eqn:expan1}),(\ref{eqn:expan15}),(\ref{eqn:expan2}) and (\ref{eqn:expan3}):
\begin{equation}
\left\{
\partial_{\tau}^{2} + \left( \eta_{\sigma}(\tau)+\frac{1}{\tau} \right) \partial_{\tau}
+ \omega_{\sigma}^{2}(\tau) + 6\lambda f_{\pi} \sigma^{(0)}(\tau)
\right\} \sigma^{(1)}(\tau,k_{T},k_{\eta}) = 0 ,
\label{eqn:appro_lambda}
\end{equation}
where 
$\sigma^{(1)}(\tau,k_{T},k_{\eta})$ is the Fourier transformation of $\sigma^{(1)}(\tau,x)$,
$k_{T}$ is the transverse momentum,  $k_{\eta}$ is the conjugate momentum of 
the 'rapidity' $\eta$ [eq.(\ref{def:variables})] and 
$\omega_{\sigma}^{2}(\tau,k_{T},k_{\eta}) = k_{T}^{2} + k_{\eta}^{2}/\tau^{2} + m_{\sigma}^{2}$
which does not depend on $k_{\eta}$ for large $\tau$.
$\sigma^{(1)}(\tau,k_{T},k_{\eta})  = f^{(0)}(\tau) u_{\sigma}^{(1)}(\tau,k_{T},k_{\eta})$ is 
substituted into eq.(\ref{eqn:appro_lambda})
as eq.(\ref{eqn:div}) is substituted into eq.(\ref{eqn:order0_sigma}) .
Note that  $f^{(0)}(\tau)$ is the same function 
as  
found in eq.(\ref{sol:f}) except for an overall constant. 
Then, the equation for $u_{\sigma}^{(1)}$ at large $\tau$ becomes 
\begin{eqnarray}
&&
\frac{d^{2} u_{\sigma}^{(1)}(\tau,k_{T},k_{\eta})}{d\tau^{2}} 
+ \left\{
\tilde{\omega}_{\sigma}^{2}(k_{T}) - \frac{1}{4} \eta_{\sigma}^{2}(\infty) 
\right. \nonumber \\ && \hspace{1cm} \left.
+ 6 \lambda f_{\pi} \sigma_{0} \tau^{-1/2} e^{-\frac{1}{2}\eta_{\sigma}(\infty)\tau}
\cos \left[
\left(m_{\sigma}^{2}-\frac{\eta_{\sigma}^{2}(\infty)}{4}\right)^{1/2} \tau + \theta
\right] 
\right\} u_{\sigma}^{(1)}(\tau,k_{T},k_{\eta}) = 0 , 
\label{eqn:difeq_sigma}
\end{eqnarray}
where $\tilde{\omega}_{\sigma}^{2}(k_{T}) \equiv \omega_{\sigma}^{2}(\infty,k_{T},k_{\eta})$.
Applying the change of variable, 
\begin{equation}
2 \xi = \left(m_{\sigma}^{2}-\frac{\eta_{\sigma}^{2}(\infty)}{4}\right)^{1/2} \tau + \theta ,
\label{def:z}
\end{equation}
to eq.(\ref{eqn:difeq_sigma}), we obtain
{
\setcounter{enumi}{\value{equation}}
\addtocounter{enumi}{1}
\setcounter{equation}{0}
\renewcommand{\theequation}{\theenumi\alph{equation}}
\begin{eqnarray}
&& 
\frac{d^{2} u_{\sigma}^{(1)}(\xi,k_{T},k_{\eta})}{d\xi^{2}} 
+ \left[ A_{\sigma}(k_{T}) - 2q_{\sigma}(\xi) \cos(2\xi) \right] u_{\sigma}^{(1)} (\xi,k_{T},k_{\eta}) =  0 ,
\label{eqn:mathiu_for_sigma}
\\ &&
A_{\sigma}(k_{T}) = 4 \left( 
\frac{\tilde{\omega}_{\sigma}^{2}(k_{T}) - \eta_{\sigma}^{2}(\infty)/4}
{m_{\sigma}^{2} - \eta_{\sigma}^{2}(\infty)/4}
\right) ,
\\ &&
q_{\sigma}(\xi) = - \left(
\frac{12 \lambda f_{\pi} \sigma_{0}}{m_{\sigma}^{2} - \eta_{\sigma}^{2}(\infty)/4}
\right) \tau^{-1/2} e^{-\frac{1}{2}\eta_{\sigma}(\infty)\tau} . 
\end{eqnarray}
The field $\sigma^{(1)}$ is obtained as follows: 
\setcounter{equation}{\value{enumi}}
}
\begin{equation}
\sigma^{(1)}(\tau,k_{T},k_{\eta}) = 
C_{\sigma} \tau^{-1/2} e^{-\frac{1}{2} \eta_{\sigma}(\infty)\tau} u_{\sigma}^{(1)}(\tau,k_{T},k_{\eta})
, 
\label{eqn:tatal_sigma}
\end{equation}
where $C_{\sigma}$ is a constant.
In the same manner, we obtain the equation for $\pi^{(1)}(\tau,k_{T},k_{\eta})$:
{
\setcounter{enumi}{\value{equation}}
\addtocounter{enumi}{1}
\setcounter{equation}{0}
\renewcommand{\theequation}{\theenumi\alph{equation}}
\begin{eqnarray}
&& 
\frac{d^{2} u_{\pi}^{(1)}(\xi,k_{T},k_{\eta})}{d\xi^{2}} + 
\left[ A_{\pi}(k_{T}) - 2q_{\pi}(\xi) \cos (2\xi) \right] u_{\pi}^{(1)}(\xi,k_{T},k_{\eta}) = 0 ,
\label{eqn:mathiu_for_pi}
\\ && 
A_{\pi}(k_{T}) = 4 \left(
\frac{\tilde{\omega}_{\pi}(k_{T}) - \eta_{\pi}^{2}(\infty)/4}
{m_{\sigma}^{2} - \eta_{\sigma}^{2}(\infty)/4}
\right),
\label{eqn:api}
\\ && 
q_{\pi}(\xi) = -\left(
\frac{4 \lambda f_{\pi} \sigma_{0}}{m_{\sigma}^{2} - \eta_{\sigma}^{2}(\infty)/4}
\right) \tau^{-1/2} e^{-\frac{1}{2}\eta_{\sigma}(\infty)\tau} ,
\label{eqn:cof_pre_cos_pi}
\end{eqnarray}
where $\tilde{\omega}_{\pi}^{2}(k_{T}) = k_{T}^{2} + m_{\pi}^{2}$.
The field $\pi^{(1)}$ is also obtained as follows: 
\setcounter{equation}{\value{enumi}}
}
\begin{equation}
\pi^{(1)}(\tau,k_{T},k_{\eta}) = 
C_{\pi} \tau^{-1/2} e^{-\frac{1}{2} \eta_{\pi}(\infty)\tau} u_{\pi}^{(1)}(\tau,k_{T},k_{\eta}), 
\label{eqn:total_pi}
\end{equation}
where $C_{\pi}$ is also a constant.
The dependence of $k_{T}$ and $k_{\eta}$ is included in only $A_{\sigma}$ and $A_{\pi}$
in eqs.(\ref{eqn:mathiu_for_sigma}) and (\ref{eqn:mathiu_for_pi}).

Eqs.(\ref{eqn:mathiu_for_sigma}) and (\ref{eqn:mathiu_for_pi}) are similar  to 
Mathieu equation which is the basic equation for the parametric amplification. 
Mathieu equation is
\begin{equation}
\frac{d^{2}F(\xi)}{d\xi^{2}} + \left[ A - 2q \cos(2\xi) \right] F(\xi) = 0 .
\label{eqn:Mathieu}
\end{equation}
The coefficient in front of cosine is a constant in Mathieu equation, 
while it is a time dependent function, i.e. a function of $\xi$,  in the present case.
The stable and unstable regions of the solution of Mathieu equation 
are described in the $q$-$A$ plane. 
The unstable regions at $q \sim 0$ correspond to $A = 1,4,9,\cdots$. 
The time dependence of $q$ in the present case do not change enhanced modes
corresponding to $A = 1,4,9,\cdots$, 
while it changes the magnitude of amplification. 
In the linear $\sigma$ model, 
the friction coefficient of $\pi$ is almost zero at zero temperature,
while that of $\sigma$ is not.  
The denominator of $A_{\sigma}$ is not negative 
for $\eta_{\sigma}(\infty) \le 2 m_{\sigma}$.
Then, the minimum of $A_{\sigma}$ is not less than 4 ,
while that of $A_{\pi}$ is smaller than 1. 
Only $\pi$ has the amplification mode for $A=1$.
The amplified modes are determined by $A=1,4,\cdots$ and has $\eta$ dependence. 
The $\eta$ dependence is weak for $\eta_{\sigma}^{2} / (4 m_{\sigma}^{2}) << 1 $.
$\eta_{\sigma}^{2} / (4 m_{\sigma}^{2})$ 
is about 0.1 for $\eta_{\sigma} = 2 {\rm fm}^{-1}$ and $m_{\sigma} = 600$MeV. 
In this study, it is apparent that the $\eta$ dependence of the amplified modes is weak 
since this condition is hold in the following calculations. 
Once the parameters of the linear $\sigma$ model and the magnitude of the friction are chosen, 
the amplified modes are calculated easily. 
As an example, $f_{\pi}$ = 92.5 MeV, $v$ = 87.4 MeV, $\lambda = 20$ are chosen for numerical calculations 
in the next section.
These parameters generate $m_{\sigma} \sim 600$ MeV, $m_{\pi} \sim 135$MeV.
The amplified modes for $\eta_{\sigma} = 0.5{\rm fm}^{-1}$ and $\eta_{\pi}=0{\rm fm}^{-1}$ are as follows.
The amplified modes $k_{T}$ for $\pi$ fields is about 267 MeV for $A_{\pi}=1$ , 583 MeV for $A_{\pi}=4$ and so on. 
That for $\sigma$ field is about 669 MeV for $A_{\sigma}=9$. 
A peak or peaks will appear in the $k_{T}$ distribution
if the parametric amplification is strong enough.
There is no mode which  corresponds to $A_{\sigma}=1$
and the mode for $A_{\sigma} = 4$ corresponds to zero mode (condensates). 
However, zero mode is the background field for the finite modes 
and $\sigma^{(1)}(x)$ does not include zero mode. 
Therefore $A_{\sigma} = 4$ cannot be a candidate for the parametric amplification.

\section{Numerical Calculation}
\label{sec:numerical}
The remaining task is to solve eqs.(\ref{eqn:mathiu_for_pi}) and 
(\ref{eqn:mathiu_for_sigma}) numerically with an adequate initial condition.
The amplitude of the oscillator is given as an initial condition.
On the other hand, we choose some magnitude of the friction referring the
results by some authors.
In subsec.\ref{subsec:inicon}, 
the initial condition and the initial momentum distribution 
to solve the equation are 
chosen to mimic the situation in high energy heavy ion collisions.
In subsec.\ref{subsec:signal}, 
we discuss the method to extract the enhancement due to parametric amplification
in the transverse momentum distribution.
In subsec.\ref{subsec:num_results},
the equation of motion is solved numerically with the initial condition 
given in subsec.\ref{subsec:inicon} and 
the characteristics of 
the enhancement formulated in  subsec.\ref{subsec:signal} is shown explicitly.

\subsection{Initial condition and magnitude of the friction}
\label{subsec:inicon}

In high energy heavy ion collisions, 
the system can be described roughly by scaling-hydrodynamics \cite{Bjorken} in the cooling stage. 
The time dependence of the temperature in the cooling stage is 
\begin{equation}
T(\tau) = T_{m} \left( \frac{\tau_{m}}{\tau} \right)^{1/3} \theta(\tau_{f}-\tau)  
\hspace{1cm} {\rm for \ \ } \tau > \tau_{m} , 
\end{equation}
where $T_{m}$ is the maximum temperature of the system,
$\tau_{m}$ is the time at which the temperature becomes maximum
and $\tau_{f}$ is the freezeout time at which the temperature becomes zero suddenly. 
If $T_{f}$ is given, $\tau_{f}$ is calculated as follows.
\begin{equation}
\tau_{f} = \tau_{m} \left( \frac{T_{m}}{T_{f}} \right)^{3} . 
\end{equation}

Though the friction is a time-dependent function in general, 
it is expected that its magnitude reaches a constant 
because the temperature becomes (almost) zero after the freezeout.
Then, we use the friction around the true minimum of the potential at zero temperature.  
The magnitude on the chirally symmetric vacuum at $T \sim  T_{c}$
obtained by Bir\'o and Greiner is $2.2 {\ \rm fm}^{-1}$ \cite{Biro}
and 
the magnitudes on chirally broken vacuum at $T=0$ 
obtained by Rischke are about $3 {\rm fm}^{-1}$  for $\sigma$ field and 0 for $\pi$ fields \cite{Rischke}. 
The contribution from nonzero modes are summed to obtain the friction for 
zero mode in such calculations.
However, the friction used in this paper comes from only hard modes. 
Its magnitude is smaller than that obtained by summing the contributions from all nonzero modes.
For these reasons, 
the following values are used in the numerical calculations:
\begin{equation}
\eta_{\sigma} = (0.25, 0.5, 1.0, 2.0) {\ \ \rm fm}^{-1} ,
\hspace{1cm}
\eta_{\pi} = 0.
\end{equation}
There are other three parameters in the linear $\sigma$ model.
We use the same values as in the previous section:  
$f_{\pi} = 92.5$ MeV, $v$ = 87.4 MeV and $\lambda=20$.

In the previous papers \cite{Ishihara2,Ishihara3}, 
it was found that the initial condition is important for
the chiral symmetry restoration in high energy heavy ion collisions.
Then, it is conceivable that the initial condition for a heavy ion collision 
affects the initial condition for the parametric amplification.
As already stated, the condensate moves along the sigma axis because it reflects 
the initial condition before the collision.
The probable initial condition for the parametric amplification in heavy ion collisions 
is 
\begin{equation}
\left. \left(\sigma^{(0)},\pi_{1}^{(0)},\pi_{2}^{(0)},\pi_{3}^{(0)} \right) 
\right|_{\tau = \tau_{f}} 
= \left( \sigma_{\rm ini} ,0,0,0 \right).
\end{equation}
The time derivatives of the fields are needed to calculate the time development. 
The condensate is almost near the minimum of the finite temperature effective 
potential in strong friction cases (of course, not too strong) and 
the velocity of the condensate is small enough. 
On the contrary, the velocity is not small in weak friction cases. 
The amplitude of the condensate in nonzero velocity cases is larger than that in zero velocity case.
The larger the amplitude of the condensate is, the stronger the amplification is.
The amplification with the initial condition that 
the time derivatives of the condensate are zero
is weakest compared with the amplification with the initial condition 
that the time derivatives are not zero. 
It will be concluded that the parametric amplification will occur 
in general if the amplification is found in zero velocity case.
The velocity of the condensate at the initial time is thus taken to be zero.
Then, 
the angle $\theta$ in eq.(\ref{eqn:difeq_sigma}) is determined by 
$\left. d\sigma^{(0)}/d\tau \right|_{\tau = \tau_{f}} = 0$ :
\begin{equation}
\theta = \arctan \left( - 
\frac{1+\eta_{\sigma}(\infty) \tau_{f}}
{2\tau_{f}\left(m_{\sigma}^{2}-\eta_{\sigma}^{2}(\infty)/4\right)^{1/2}}
\right)
- \left(m_{\sigma}^{2}-\frac{\eta_{\sigma}^{2}(\infty)}{4}\right)^{1/2} \tau_{f} . 
\label{eqn:init_velocity}
\end{equation}

The momentum distribution is proportional to the square of the fields. 
On the other hand, 
the momentum distribution depends on only $k_{T}$ because of the scaling property.
Therefore, the field which is the square root of the momentum distribution depends on only $k_{T}$.
Then, we take the initial value of 
$\sigma^{(1)}(\tau_{f};k_{T},k_{\eta})$ and $\pi^{(1)}(\tau_{f};k_{T},k_{\eta})$ as 
\begin{equation}
\sigma^{(1)}(\tau_{f};k_{T},k_{\eta}) \propto \sqrt{g(\vec{k}_{T};m_{\sigma})} 
, \hspace{1cm}
\pi^{(1)}(\tau_{f};k_{T},k_{\eta}) \propto \sqrt{g(\vec{k}_{T};m_{\pi})} .
\label{field:initial_cond}
\end{equation}
where $g(\vec{k}_{T};m)$ is the initial distribution function.
The absolute magnitude which is not given in eq.(\ref{field:initial_cond})
is unimportant in the following discussion
because we will consider only the relative ratio. 
If the system is locally thermalized at the initial time, 
$g(\vec{k}_{T};m)$ is chosen as follows:
\begin{equation}
g(\vec{k}_{T};m) = \frac{1}{\exp \left(\frac{\sqrt{m^{2}+ \vec{k}_{T}^{2}}}{T_{f}}\right) - 1}.
\label{eqn:thermal}
\end{equation}

\subsection{Extraction of the peaks due to parametric amplification}
\label{subsec:signal}
One important problem is how to extract the peaks in the transverse momentum distribution
generated by the parametric amplification in numerical calculations.
The amplitude $\sigma$ and $\pi$ become zero asymptotically for all mode 
because of the expansion and the dissipation. 
First, we introduce the following quantities to extract the magnitude of the amplification 
of the fields.
\begin{equation}
R_{\alpha}(\tau; A_{\alpha}, \tau_{f}) \equiv
R_{\alpha}(\tau; k_{T}, k_{\eta}, \tau_{f}) 
\stackrel{\rm def}{=}
\frac{\psi_{\alpha}^{(1)}(\tau; k_{T}, k_{\eta})}{\psi_{\alpha}^{(1)}(\tau_{f}; k_{T}, k_{\eta})} ,
\end{equation}
where $\alpha = \pi, \sigma$ and  
$\psi_{\pi}^{(1)} = \pi^{(1)}$, $\psi_{\sigma}^{(1)} = \sigma^{(1)}$ .
The ratio $R_{\alpha}$ oscillates as a function of $\tau$. 
The amplification of the fields is seen explicitly by removing the damping factor,
$\exp[-\eta_{\sigma}(\infty) \tau /2] \tau^{-1/2}$ from $R_{\alpha}$.
It is performed by using the following quantity:
\begin{equation}
r_{\alpha}(\tau; A_{\alpha}, \tau_{f}) \stackrel{\rm def}{=}
 u^{(1)}_{\alpha}(\tau, k_{T}) / u^{(1)}_{\alpha}(\tau_{f}, k_{T}).
\end{equation}

It is useful to consider the envelop of $R_{\alpha}$ and $r_{\alpha}$
for our purpose.
The envelope of $r_{\alpha}$ is denoted by $r^{e}_{\alpha}$.
However,  $R_{\alpha}$ is not an adequate quantity to extract the amplification 
and the envelop of $R_{\alpha}$ too because of the damping. 
Fortunately, the damping factor is momentum independent. 
The reduction of the amplitude by the expansion and the dissipation are cancelled 
by taking the ratio between different modes.
We use the $r^{e}_{\alpha}$  
because the envelop is easily calculated for $r_{\alpha}$. 
\begin{equation} 
P_{\alpha}(A_{\alpha},A_{\alpha,r}) \stackrel{\rm def}{=} 
\lim_{\tau \rightarrow \infty}
\frac{r^{e}_{\alpha}(\tau; A_{\alpha}, \tau_{f})}
{r^{e}_{\alpha}(\tau; A_{\alpha,r}, \tau_{f})} 
,
\end{equation}
where $A_{\pi, r}$ and $A_{\sigma, r}$ are the reference to other modes, $A_{\pi}$
and $A_{\sigma}$ respectively.
Since the ratio of the momentum distribution is proportional to $\left[ P(A,A_{r}) \right]^{2}$, 
it links directly the amplification and the transverse momentum distribution. 
The effects of the initial distribution are taken into account by multiplying 
the square root of the initial function as follows:
\begin{equation}
P^{\rm new}(A,A_{r}) \stackrel{\rm def}{=} P(A,A_{r}) \sqrt{\frac{\tilde{g}(A)}{\tilde{g}(A_{r})}},
\hspace{1cm}
\tilde{g}(A) = g(k_{T}),
\label{def:Pnew}
\end{equation}
where $g(k_{T})$ is the initial distribution function introduced in eq.(\ref{eqn:thermal}).
The transverse momentum distribution can be calculated by eq.(\ref{def:Pnew})
for various initial distributions.

The time derivative terms  are not explicitly included in the quantities introduced in 
this subsection because the effect of these terms are small. 
These effects in terms of above quantities are evaluated in the appendix \ref{appendix}.

\subsection{Numerical Results}
\label{subsec:num_results}
What we would like to know are 
1) whether parametric amplification occurs or not, and 
2) whether its signal can be observed experimentally or not.
The answer to the question 1) can be obtained by studying $r_{\alpha}$.
Hereafter, we consider only the $\pi$ fields 
because $\sigma$ field will not be amplified strongly.
The reasons that the sigma field is not amplified strongly are that
1) the smallest amplified mode corresponds to $A_{\sigma} =9$ ,
and 
2) the effect of the friction  for $\sigma$ [eq.(\ref{eqn:tatal_sigma})] does not vanish, 
while that for $\pi$ vanishes [eq.(\ref{eqn:total_pi})] at zero temperature.

First, we show the time development of $R_{\pi}$ and $r_{\pi}$
for $A_{\pi} = 1$, $\sigma_{\rm ini} = - 30 {\rm MeV}$, $\eta_{\sigma}=0.25{\rm fm}^{-1}$
and $T_{f} = 100$MeV.
The evolution of $R_{\pi}$ is shown in Fig.\ref{fig:Rratio} 
and that of $r_{\pi}$ is shown in Fig.\ref{fig:uratio}.
The amplitude of $R_{\pi}$ increases temporarily because of  the parametric amplification 
and then decreases because of the expansion and the dissipation.
Removing the effect of the expansion and the dissipation, 
we can see the amplification obviously shown in Fig.\ref{fig:uratio}.  
The amplitude of $r_{\pi}$ becomes a constant 
since the factor $q_{\pi}(\xi)$ [see eq.(\ref{eqn:cof_pre_cos_pi})] becomes zero.

The asymptotic value of the amplitude, $r_{\pi}^{e}(\tau=\infty,A_{\pi})$,
is shown in Fig.\ref{fig:Tmdep2} for various $T_{m}$ as a function of $A_{\pi}$
for $\sigma_{\rm ini} = - 30 {\rm MeV}$, $\eta_{\sigma}=0.25{\rm fm}^{-1}$ and 
$T_{f} = 100$MeV.
There is a peak near $A_{\pi} \sim 1$ caused by  
the parametric amplification. 
This strong amplification may give the possibility of the observation
even when the initial distribution does not have enough amplitude at $A_{\pi} \sim 1$.
It seems that 
the amplification saturates 
as $T_{m}$ increases.
This is understood as follows. 
The increase of the amplitude by the parametric amplification depends on 
the initial amplitude of the condensates and the time interval $(s = \tau - \tau_{f})$. 
Since the amplitude at the initial time is fixed in this calculation,
the time interval is investigated here. 
Amplification described in Mathieu equation is 
characterized on the $q$-$A$ plane,  where $q$ and $A$ appear in eq.(\ref{eqn:Mathieu}).
The field is amplified if the set of the parameters $q$ and $A$ is in a unstable region 
on the $q$-$A$ plane.
The set $(q,A)$ at the initial time is shown by dot in Fig.\ref{fig:qAplane}.
The coefficient $q(\xi)$ defined by eq.(\ref{eqn:cof_pre_cos_pi})
is a time dependent function and it moves to zero as shown in Fig.\ref{fig:qAplane}, 
while it is a constant in Mathieu equation. 
The point on $q$-$A$ plane goes out of the unstable region as time goes by.
This implies that 
the smallest $q(\xi)$ for each modes exists for the field to be amplified.
Then, we consider a certain time $\tau_{Q}$ at which 
$q_{\pi}(\xi)$ in eq.(\ref{eqn:cof_pre_cos_pi}) becomes a certain value, $Q$.
We consider also $Q_{f}(\tau_{f})$ which is $q(\xi)$ when $\tau = \tau_{f}$.
With the help of eq.(\ref{eqn:cof_pre_cos_pi}), we obtain 
\begin{equation}
\exp \left( - \eta_{\sigma}(\infty) s \right) = 
\frac{Q^{2}}{Q^{2}_{f}(\tau_{f})} \left( 1 +  \frac{s}{\tau_{f}} \right), 
\label{eqn:limit}
\end{equation}
where $s = \tau_{Q} - \tau_{f}$.
The left hand side is a function of only $s$, 
while  the right hand side depends on both $s$ and $\tau_{f}$.
That $T_{m}$ goes to infinity corresponds to that  $\tau_{f}$ goes to infinity.
Then, $\tau_{f} \rightarrow \infty$ limit can be taken with a fixed $s$.
Since $Q_{f}(\tau_{f})$ converges at a certain value $Q_{f}$ 
which is obtained from eqs.(\ref{eqn:init_sigma0}), (\ref{eqn:api}) 
and (\ref{eqn:init_velocity}),
the right hand side of eq.(\ref{eqn:limit}) becomes $Q^{2}/Q_{f}^{2}$.
Then, $s$ is finite as $T_{m}$ becomes infinity, 
which is obtained by solving eq.(\ref{eqn:limit}):
\begin{equation}
s = - \frac{2}{\eta_{\sigma}(\infty)} \ln \left( Q/Q_{f} \right).
\end{equation}
This implies that the time interval for the amplification is still a finite constant
even when $T_{m}$ is infinite.
We conclude that the amplification saturates as $T_{m}$ becomes infinite
because the initial amplitude is finite and the time interval converges.

The parameters, $\eta_{\sigma}(\infty)$,  $\sigma_{\rm ini}$ and  $T_{f}$,
are varied to see the sensitivity of the amplification in the following calculations. 

The ratio of the amplifications, $r_{\pi}^{e}(\tau=\infty,A_{\pi})$,
 is shown in Fig.\ref{fig:FricDep} 
for various $\eta_{\sigma}(\infty)$ as a function of $A_{\pi}$.
The magnitude of the amplification decreases as the magnitude of the friction increases. 
It is reasonable because the oscillation of the condensate are reduced rapidly 
when the friction is strong. 
It is apparent that the amplitude of the oscillation decreases 
since $2 q_{\alpha}(\xi)$ 
in eqs.(\ref{eqn:mathiu_for_pi}) and (\ref{eqn:mathiu_for_sigma}) vanishes quickly
in the strong friction cases. 
It was found that the parametric amplification cannot occur for frictions 
which is larger than $2{\rm fm}^{-1}$. 

The ratio of amplifications , $r_{\pi}^{e}(\tau=\infty,A_{\pi})$, 
is shown in Fig.\ref{fig:SigmaIni} for 
various  $\sigma_{\rm ini}$ as a function of $A_{\pi}$.
The dependence on $\sigma_{\rm ini}$ is important for the parametric amplification 
in the presence of the expansion and the friction 
because the amplitude of the oscillator (condensate) decreases as time goes by. 
However, we can find a peak near $A_{\pi} \sim 1$, at least, for $\sigma_{\rm ini} \le -10$ MeV.
Then, it may be possible to find this peak in the transverse momentum distribution.

The ratio of amplification , $r_{\pi}^{e}(\tau=\infty,A_{\pi})$, 
is shown in Fig.\ref{fig:Tf} for various $T_{f}$ 
as a function of $A_{\pi}$.
It is obvious that $T_{f}$ dependence is weak.

Hereafter, 
the $A_{\pi}$ distribution, namely the transverse momentum distribution, 
is displayed taking into account the initial transverse momentum distribution.
The thermal distribution [eq.(\ref{eqn:thermal})] is assumed 
and eq.(\ref{def:Pnew}) is used.
We take $A_{r}=0.5$ which is the reference in $P^{\rm new}$ defined in eq.(\ref{def:Pnew}). 

The ratio between different modes including the effect of the initial thermal distribution, 
$P_{\pi}^{\rm new}(A_{\pi},A_{\pi,r}=0.5)$, for various $T_{m}$ 
is shown in Fig.\ref{fig:Pnew_ini-20.0_Tf100_ETA0.5_pi}.
The parameters are 
$\sigma_{\rm ini} = -20$MeV, $T_{f} = 100$MeV and $\eta=0.5{\rm fm}^{-1}$. 
The ratio between the momentum distributions corresponds to $|P_{\pi}^{\rm new}|^{2}$. 
Then, it is found from Fig.\ref{fig:Pnew_ini-20.0_Tf100_ETA0.5_pi} that 
the ratio at $A_{\pi} \sim 1$ is 4 times larger than  that at $A_{\pi} =0.5$ 
for $T_{m}=300$MeV.
The increase in small $A_{\pi}$ and the decrease in large $A_{\pi}$ reflect 
the initial thermal distribution.
Then, $P^{\rm new}(A,A_{r})$ for large $A$ limit becomes zero 
when the initial distribution is thermal.
Similar calculations can be performed to obtain $P^{\rm new}_{\pi}$ for various parameters. 
For example, $P^{\rm new}_{\pi}(A_{\pi} \sim 1, A_{\pi,r}=0.5) \sim 12.7$ 
for $\sigma_{\rm ini} = -30.0$MeV, $T_{m}= 300$MeV, $T_{f} = 100$MeV and 
$\eta_{\sigma} = 0.25{\rm fm}^{-1}$. 

There are many $A$'s at which a field is amplified in Mathieu equation. 
The amplification around $A_{\pi}=4$ is also investigated,  
but strong amplification is not found.
This comes from the property of  Mathieu equation and 
the small initial amplitude in the initial thermal distribution compared with that 
at $A_{\pi} \sim 1$.
Then, it is difficult to observe the effects of amplification except for $A_{\pi} \sim 1$.

\section{Conclusions and discussions}
\label{sec:conclude}
We investigate the effects of the expansion and the friction 
on the parametric amplification in one dimensional scaling case 
within the linear $\sigma$ model.
The equation of motion at zero temperature is derived 
in the case where the sigma condensate oscillates around the minimum of the potential.
From this study, 
we found that the amplitude of the oscillator (zero mode) and the magnitude of the friction 
are essentially important for the parametric amplification in one dimensional scaling case.
In a weak friction case,
the strong enhancement will occur at $A_{\pi} \sim 1$. ($k_{T} \sim 267$MeV) 
in the pion transverse momentum distribution and it may be possible to be observed 
by the normalized distribution, $\left| P^{\rm new} \right|^{2}$. 
Then, we have a chance to catch the signal of the {\it temporal} restoration 
of chiral symmetry \cite{Ishihara2}.
Contrary to this, 
it may be difficult for the field to be amplified by the 
parametric amplification in a strong friction case.
On the other hand, high maximum temperature and/or sufficiently strong friction 
are needed for chiral symmetry to be restored for a long time 
as we found before \cite{Ishihara3}.

High maximum temperatures (above 100 MeV) was also assumed in this study, 
which settles the condensate at an unstable point on the potential.
However, the mechanism of parametric amplification does not always 
require high temperature. 
The {\it partial} chiral symmetry restoration may be also observed  
if the amplitude of the condensate is large enough.

It is possible to devide the modes of the field  into three parts, 
the condensate, soft modes and hard modes. 
(See Fig.\ref{threemodes}) 
The hard modes can be regarded as 
the cause of the friction for the condensate and soft modes.
There are the energy flows from the condensate and soft modes to hard modes. 
In the same way, 
there is the energy flow from the condensate to some soft modes. 
The decreasing of the zero mode amplitude comes from 
the energy dissipation due to the friction and the effect of the expansion.
However, the energy flow from the condensate to the soft modes 
is not taken into account in the present study. 
Back reaction is discarded too. 
They will be taken into account in the future study.

\begin{acknowledgments}
I would like to thank F. Takagi for a number of helpful suggestions.
\end{acknowledgments}

\appendix
\section{The effects of the time derivative terms}
\label{appendix}
In this appendix, 
the effect of the time derivative terms of the fields is checked 
for the quantities introduced in sec.\ref{subsec:signal}.
First, we consider the effect on $R_{\alpha}$.
The ratio of the number-density like quantities is evaluated in order to make it clear
that the effect is small in the present case:
\begin{equation} 
\tilde{R}_{\alpha}^{(2)}(\tau; A_{\alpha}, \tau_{f}) = 
\frac{\left[ \psi^{(1)}_{\alpha}(\tau; A_{\alpha}) \right]^{2}
    + {\tilde{\omega}_{\alpha}^{-2}} \left[ {d \psi^{(1)}_{\alpha}}/{d\tau} \right]^{2}  }
{\left[ \psi^{(1)}_{\alpha} (\tau_{f}; A_{\alpha}) \right]^{2}
    + {\tilde{\omega}_{\alpha}^{-2}} \left[ 
      \left. {d \psi^{(1)}_{\alpha}}/{d\tau} \right|_{\tau=\tau_{f}}
    \right]^{2} }. 
\end{equation}
Note that 
$\tilde{\omega}_{\alpha}$ does not have $k_{\eta}^{2}/\tau^{2}$ term.
It is ignored because $\tau$ is large enough. 
The time derivative of  $\psi^{(1)}_{\alpha}$ is
\begin{equation}
\frac{d \psi^{(1)}_{\alpha}}{d\tau} = 
  \frac{\exp \left(- \eta_{\alpha} \tau /2 \right)}{\tau^{1/2}} \frac{d u_{\alpha}}{d \tau} 
- \frac{\eta_{\alpha}}{2}   \frac{\exp \left(- \eta_{\alpha} \tau /2 \right)}{\tau^{1/2}} 
u_{\alpha}
- \frac{1}{2}   \frac{\exp \left(- \eta_{\alpha} \tau /2 \right)}{\tau^{3/2}} u_{\alpha}
,
\end{equation}
where $\eta_{\alpha}$ is the friction for $\psi^{(1)}_{\alpha}$.
At the time ($\tau_{e}$) at which $du_{\alpha}/d\tau = 0$, one has 
\begin{equation}
\left. \frac{d \psi^{(1)}_{\alpha}}{d\tau}\right|_{\tau = \tau_{e}} = 
- \left( \frac{\eta_{\alpha}}{2} + \frac{1}{2 \tau_{e}} \right)
\frac{\exp \left(- \eta_{\alpha} \tau_{e} /2 \right)}{\left( \tau_{e} \right)^{1/2}} u_{\alpha}.
\end{equation}
The same expression is obtained at the initial time 
for the intial condition, $\left( du_{\alpha}/d\tau \right)_{\tau = \tau_{f}} = 0$. 
$\tilde{R}_{\alpha}^{(2)}$ at $\tau_{e}$ is evaluated 
since the amplification is evaluated at the envelop of $r_{\alpha}$.
\begin{equation}
\tilde{R}_{\alpha}^{(2)}(\tau_{e}) = 
\frac{\exp \left[ - \eta_{\alpha} \left( \tau_{e} - \tau_{f} \right) \right]}{\tau_{e} / \tau_{f}} 
\left[
\frac{1 + \left(2 \tilde{\omega}_{\alpha}\right)^{-2}\left( \eta_{\alpha} + \tau_{e}^{-1} \right)^{2} }
{1 + \left( 2 \tilde{\omega}_{\alpha} \right) ^{-2} \left( \eta_{\alpha} + \tau_{f}^{-1} \right)^{2} }
\right]
r_{\alpha}^{2}
\stackrel{\rm def}{=} 
\frac{\exp \left[ - \eta_{\alpha} \left( \tau_{e} - \tau_{f} \right) \right]}{\tau_{e} / \tau_{f}} 
\ D_{\alpha}(\tau_{e},\tau_{f},\eta) \ r_{\alpha}^{2}.
\end{equation}
The effect of the time derivative terms is included in $D_{\alpha}(\tau_{e},\tau_{f},\eta)$. 
Apparently,  
$D_{\alpha}(\infty,\tau_{f},\eta_{\alpha}) \le D_{\alpha}(\tau_{e},\tau_{f},\eta_{\alpha}) \le 
D_{\alpha}(\tau_{f},\tau_{f},\eta_{\alpha}) = 1 $.
Then 
$D_{\pi}(\infty,\tau_{f},0)$ for $\pi^{(1)}$ and 
the minimum of $D_{\sigma}(\infty,\tau_{f},\eta_{\sigma})$ 
with respect to $\eta_{\sigma}$ for $\sigma^{(1)}$ 
are evaluated with a fixed $\tau_{f}$:
{
\setcounter{enumi}{\value{equation}}
\addtocounter{enumi}{1}
\setcounter{equation}{0}
\renewcommand{\theequation}{\theenumi\alph{equation}}
\begin{eqnarray}
D_{\pi}(\infty,\tau_{f},0) &=& 
\left[ 1 + 1 / \left( 4 \tau_{f}^{2} \tilde{\omega}_{\pi}^{2} \right) \right]^{-1} ,
\\
\left. D_{\sigma}(\infty,\tau_{f},\eta_{\sigma}) \right|_{\rm min} &=&
\frac{\sqrt{16 \tilde{\omega}_{\sigma}^{2} \tau_{f}^{2} + 1 } - 1}{\sqrt{16 \tilde{\omega}_{\sigma}^{2} \tau_{f}^{2} + 1 } + 1}.
\end{eqnarray}
$D_{\alpha}$ at $k_{T}=0$ is the minimum on the $k_{T}$ axis.
$\tau_{f}$ is some 10 fm, $m_{\pi}$ is 135 MeV and $m_{\sigma}$ = 600 MeV in the present case.
In such a case, 
$D_{\pi}$ and $D_{\sigma}$ at the minimum  are 
about 0.9947 and is about 0.9837 respectively
for $\tau_{f} = 10$ fm and $k_{T}=0$
The maximum difference between $D_{\alpha}$ 
with time derivative terms and without is less than 2\%.
These are small enough.

The effect of the time derivative terms for $P_{\alpha}(A_{\alpha},A_{\alpha,r})$ is evaluated
in the same way. 
$P_{\alpha}(A_{\alpha},A_{\alpha,r})$ is the ratio between different modes at the same time,
while $\tilde{R}^{(2)}_{\alpha}$ is the ratio between different times at the same mode. 
It is shown that time derivative terms do not affect on  $P_{\pi}$, 
while it is shown that time derivative terms affect slightly on $P_{\sigma}$.

{
\setcounter{enumi}{\value{figure}}
\addtocounter{enumi}{1}
\setcounter{figure}{0}
\renewcommand{\thefigure}{\theenumi\alph{figure}}
\begin{figure}
\caption{
Time development of the ratio $R_{\pi}$ with  
$A_{\pi} = 1$, $\sigma_{\rm ini} = -30$MeV, $T_{m} = 300$MeV, 
$T_{f}=100$MeV and $\eta_{\sigma}(\infty) = 0.25{\rm fm}^{-1}$.
}
\label{fig:Rratio}
\end{figure}

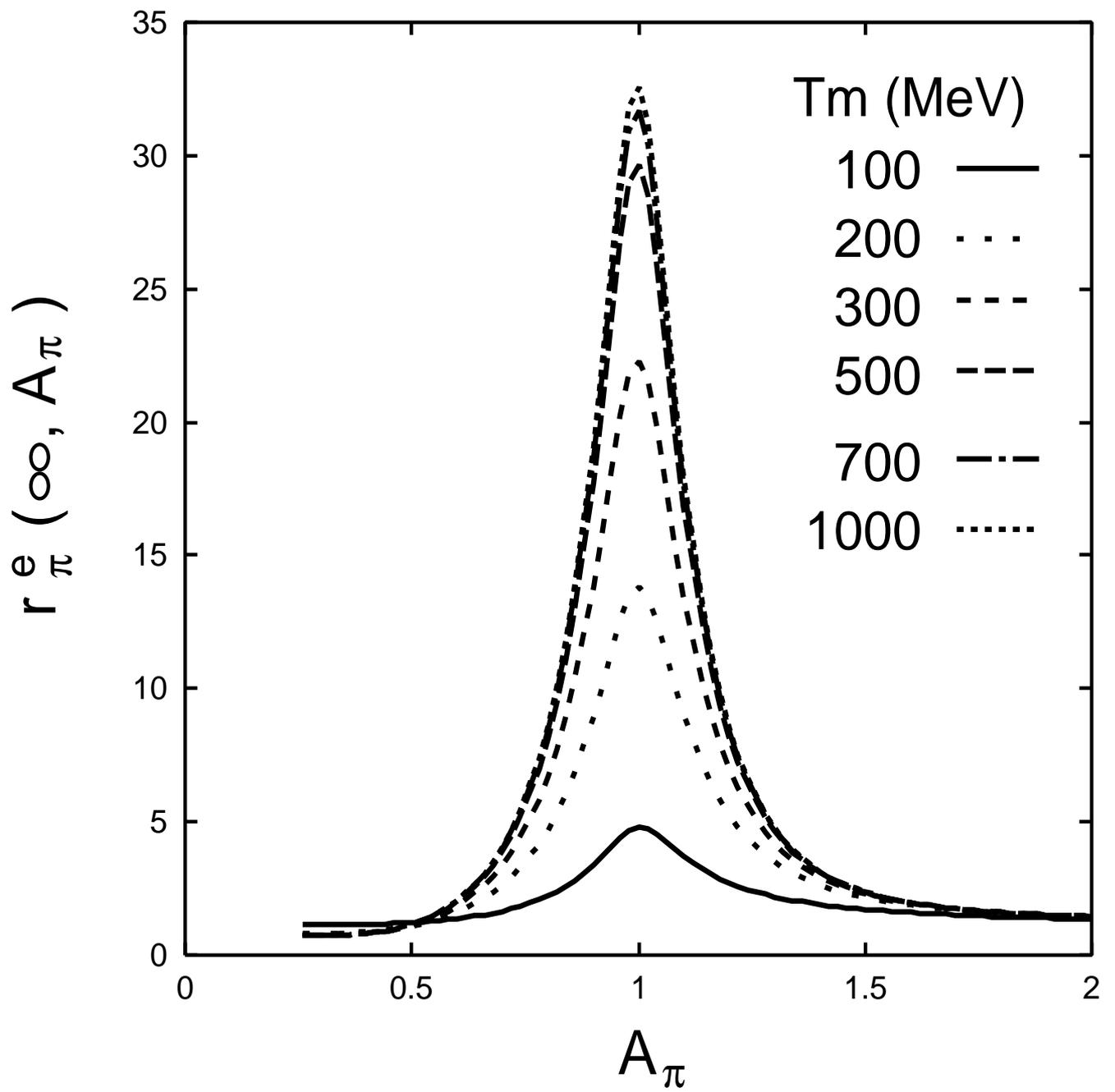
\begin{figure}
\caption{
Time development of the ratio $u_{\pi}$ with  
$A_{\pi} = 1$, $\sigma_{\rm ini} = -30$MeV, $T_{m} = 300$MeV, 
$T_{f}=100$MeV and $\eta_{\sigma}(\infty) = 0.25{\rm fm}^{-1}$.
}
\label{fig:uratio}
\end{figure}
\setcounter{figure}{\value{enumi}}
}


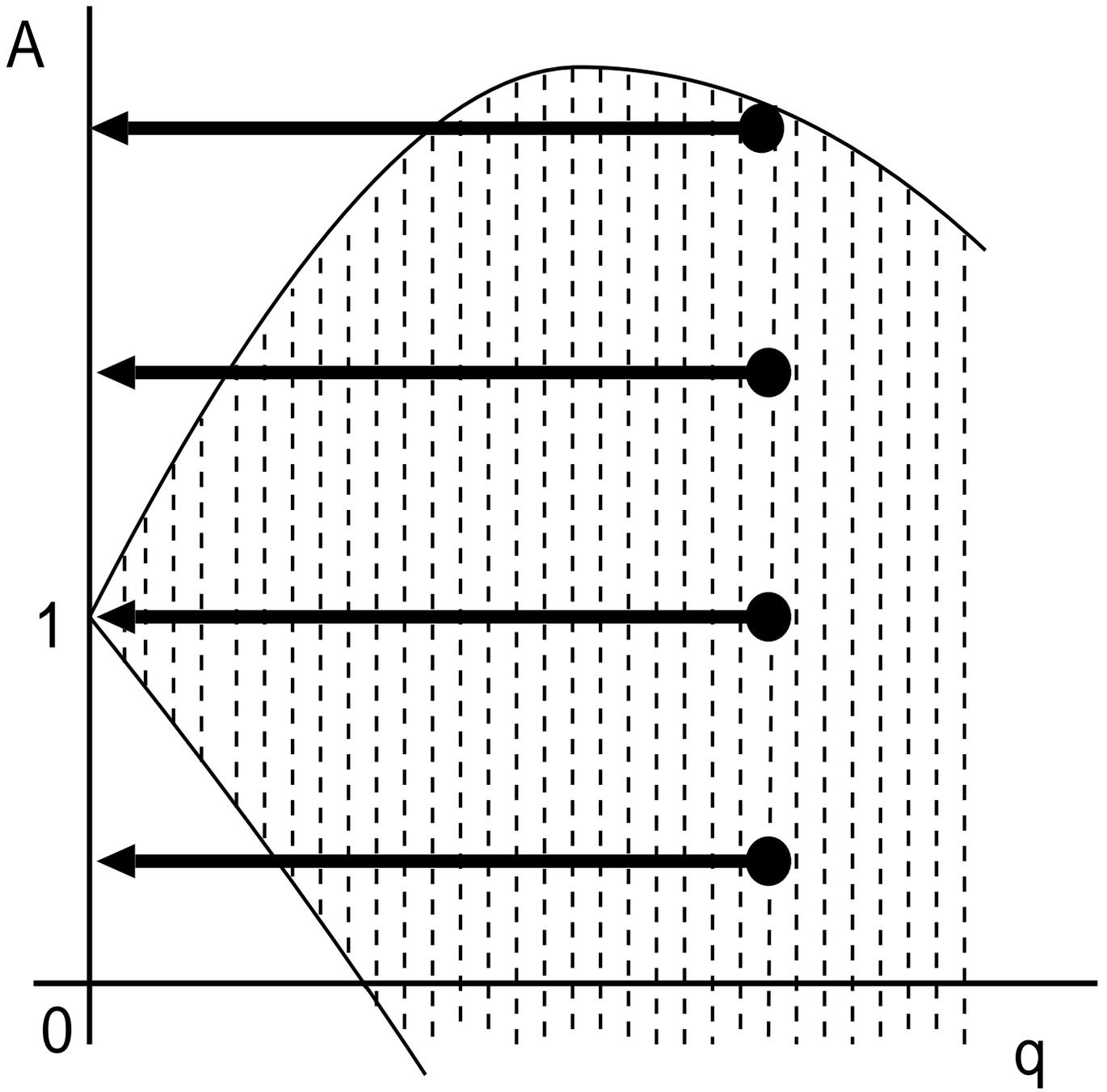
\begin{figure}
\caption{
The amplifications of the field amplitude in the broad range $T_{m}$.
The initial condition is $\sigma_{\rm ini} = -30$MeV, $\eta_{\sigma}=0.25{\rm fm}^{-1}$
and $T_{f}=100$MeV. 
}
\label{fig:Tmdep2}
\end{figure}

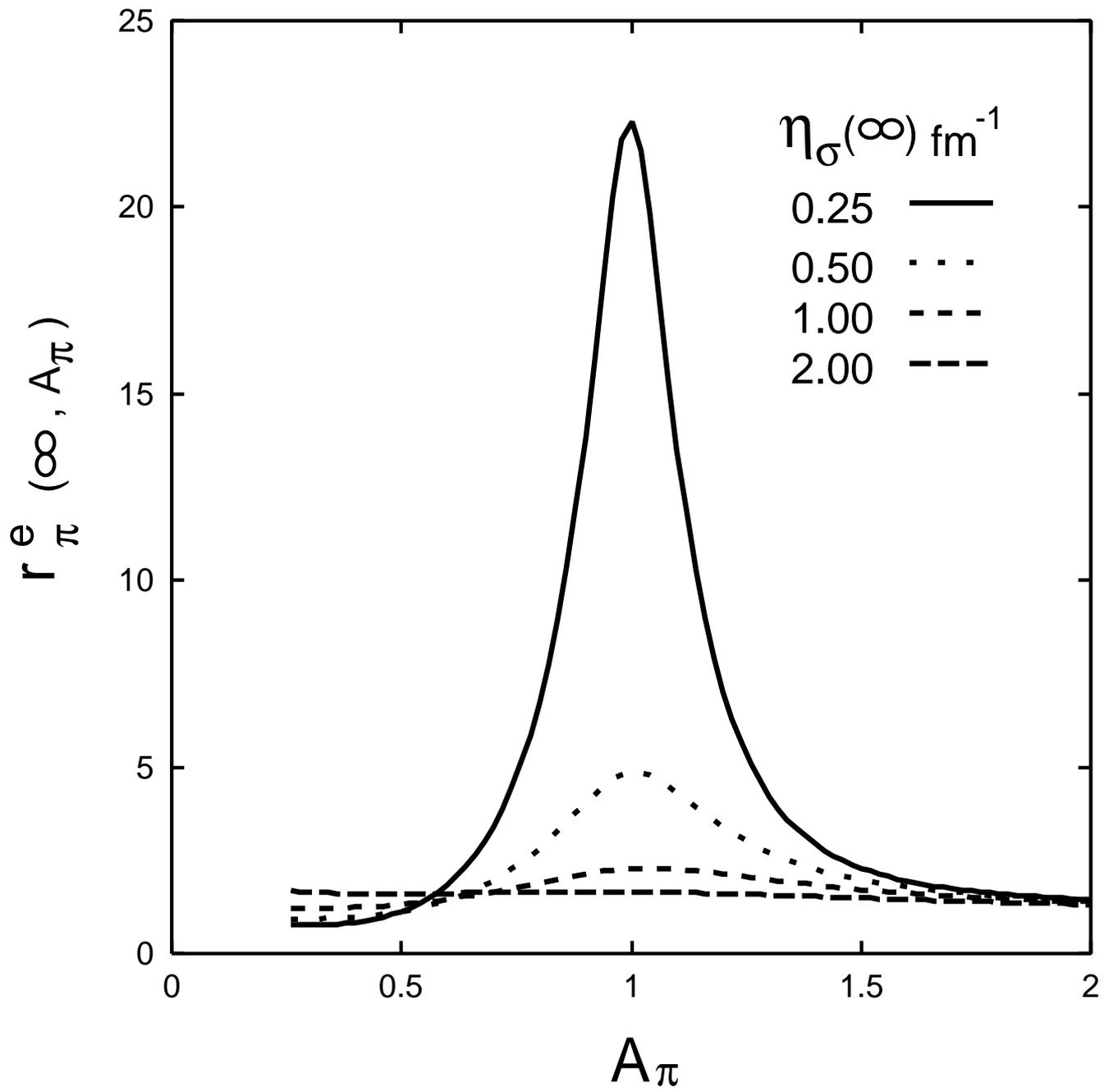
\begin{figure}
\caption{
Sketch of the stable and unstable regions around $A=1$ in the $q$-$A$ plane. 
The shaded region is corresponding to the unstable region.
The initial values in the $q$-$A$ plane is indicated by black balls.
The set $(q,A)$ moves on the thick line to the $q=0$ direction 
with a constant $A$.
}
\label{fig:qAplane}
\end{figure}


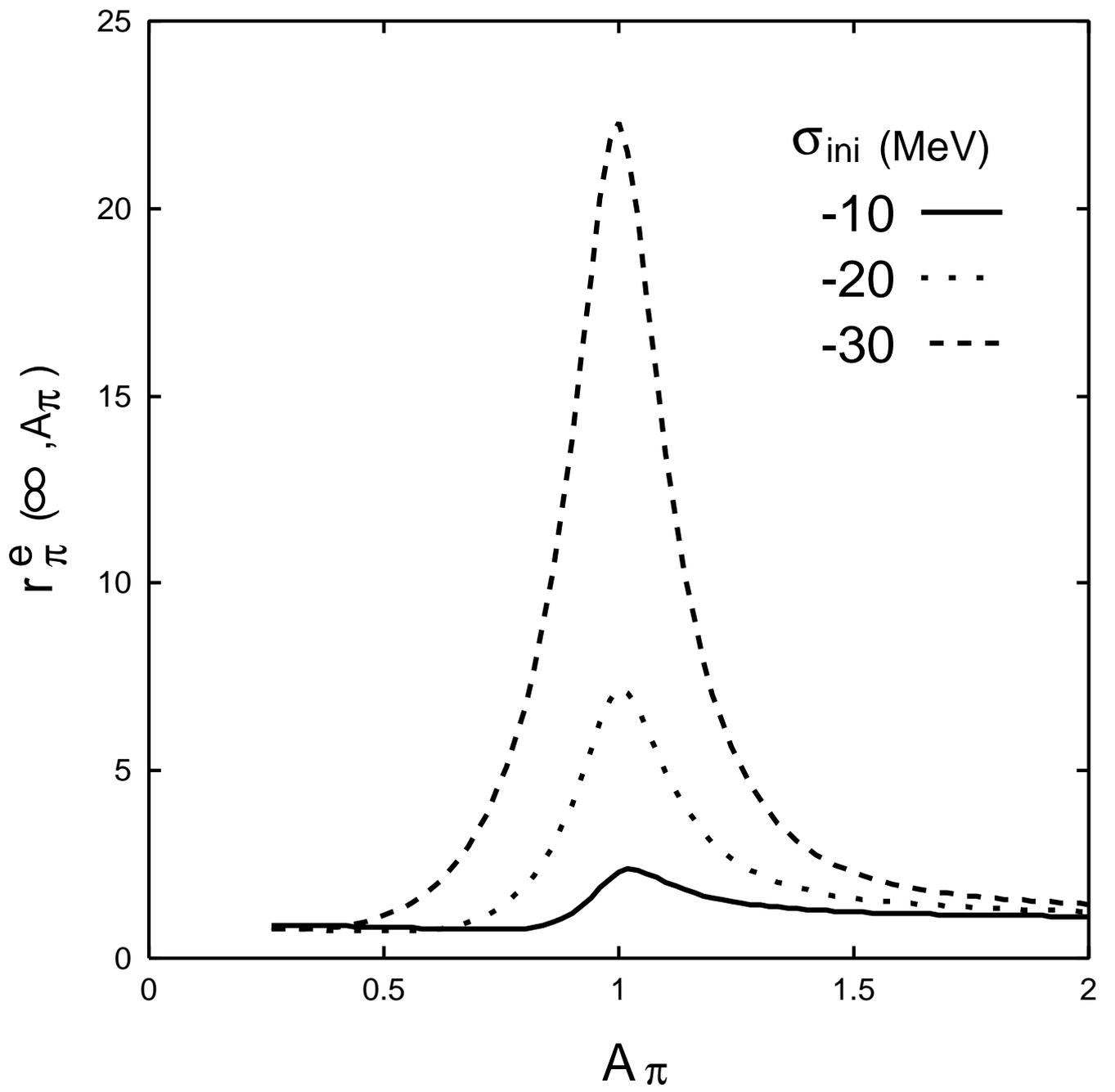
\begin{figure}
\caption{
The amplifications of the field amplitude for various magnitudes of friction.
The initial condition is $T_{m} = 300$MeV, $\sigma_{\rm ini} = -30$MeV 
and $T_{f}=100$MeV. 
}
\label{fig:FricDep}
\end{figure}

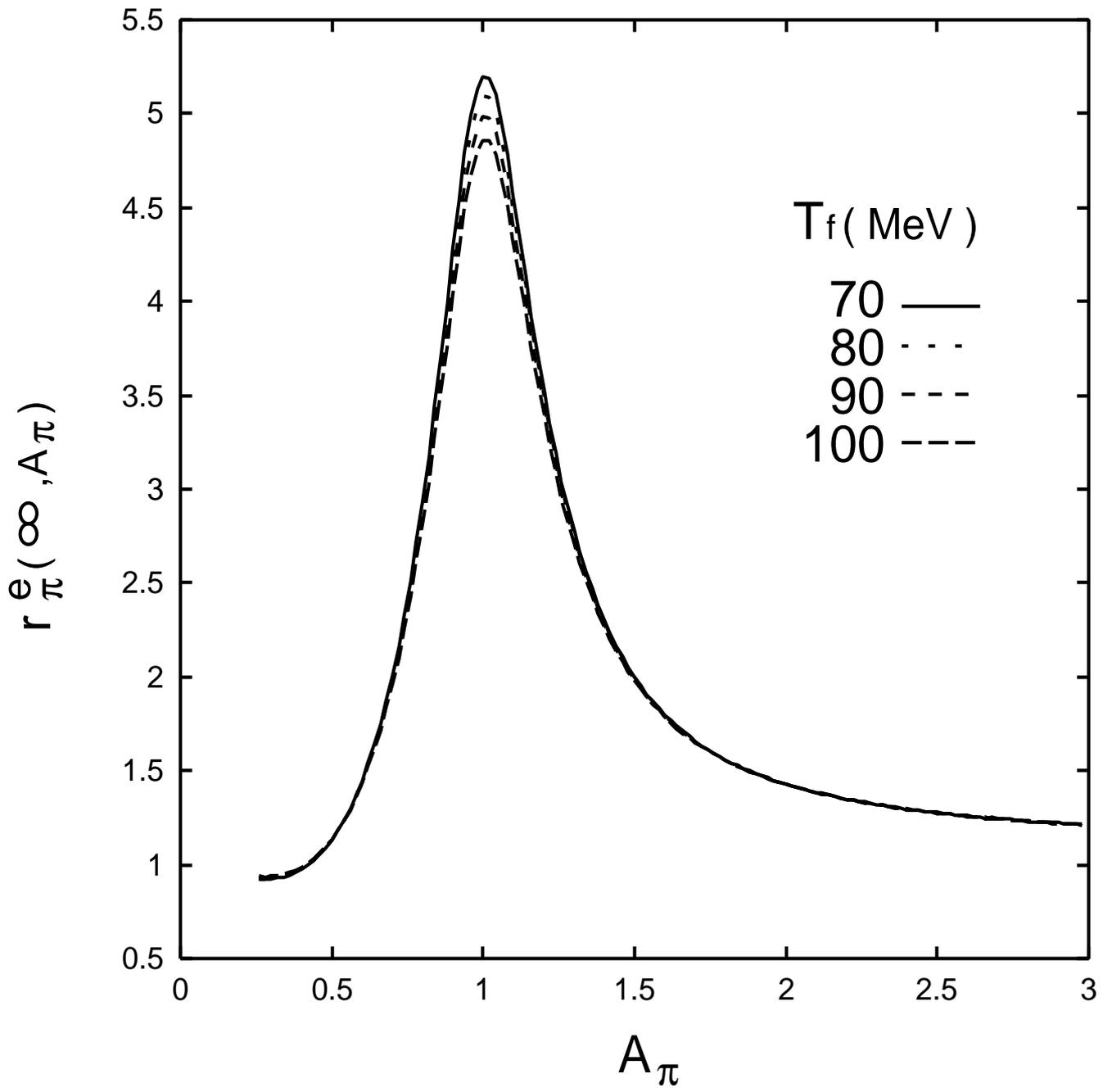
\begin{figure}
\caption{
The amplifications of the field amplitude for various $\sigma_{\rm ini}$.
The initial condition is $T_{m} = 300$MeV, $\eta_{\sigma}=0.25{\rm fm}^{-1}$
and $T_{f}=100$MeV. 
}
\label{fig:SigmaIni}
\end{figure}

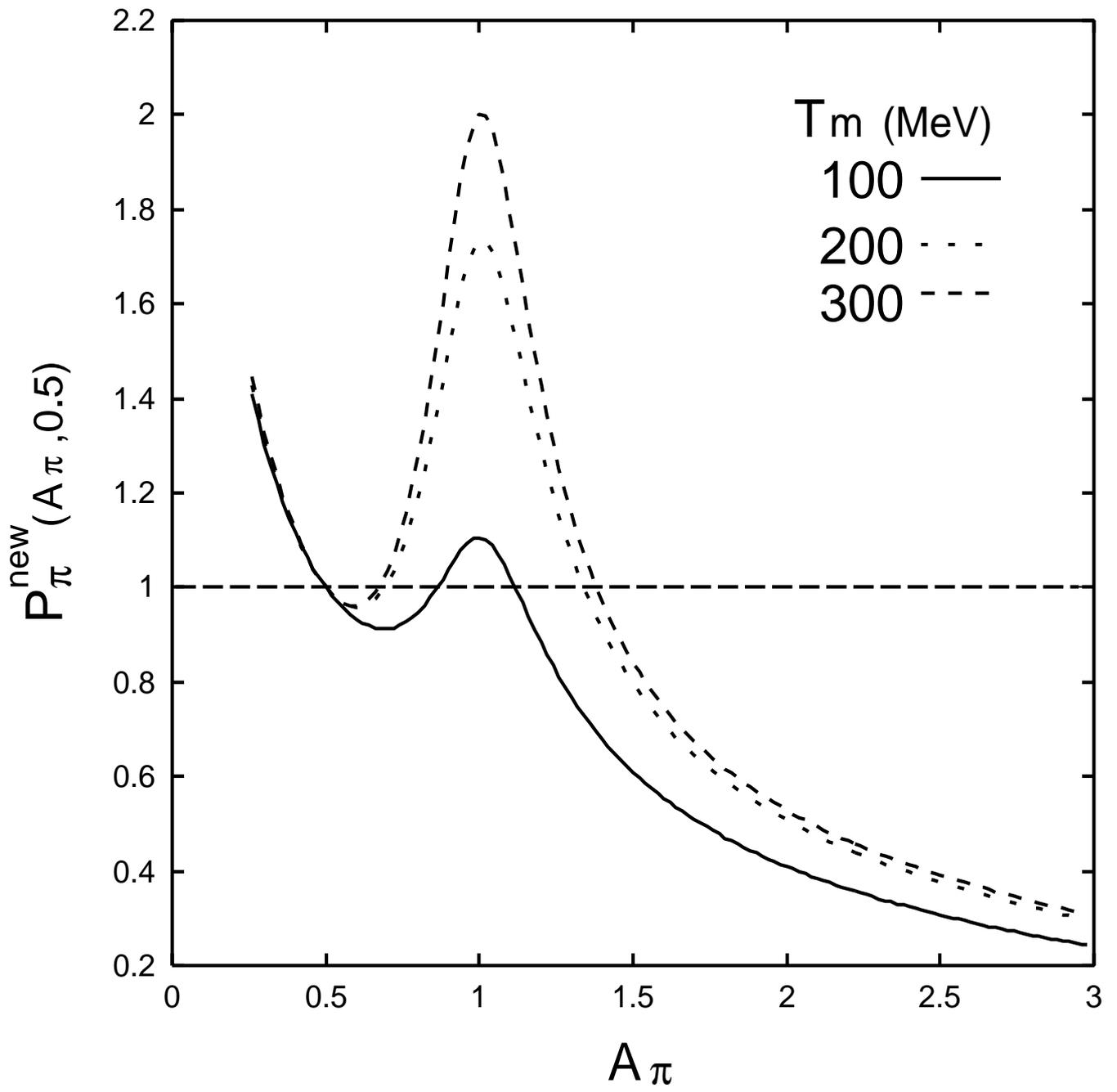
\begin{figure}
\caption{
The amplifications of the field amplitude for various $T_{f}$.
The initial condition is $T_{m} = 300$MeV, $\eta_{\sigma}=0.5{\rm fm}^{-1}$
and $\sigma_{\rm ini} = -30$MeV.
}
\label{fig:Tf}
\end{figure}

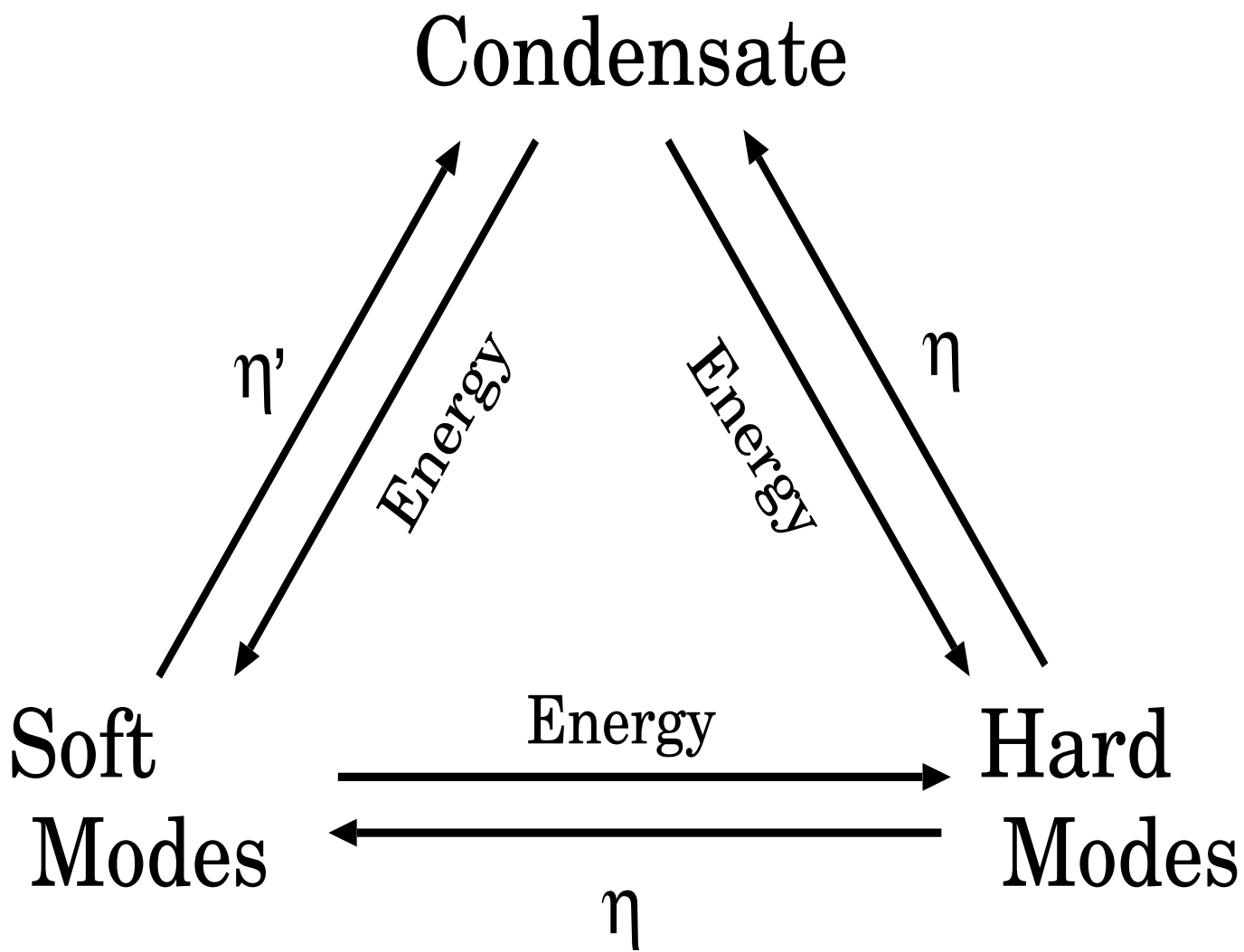
\begin{figure}
\caption{
The ratio of amplification of the amplitude for various $T_{m}$ with 
$\sigma_{\rm ini} = -20$MeV, $T_{f} = 100$MeV and $\eta=0.5{\rm fm}^{-1}$. 
}
\label{fig:Pnew_ini-20.0_Tf100_ETA0.5_pi}
\end{figure}

\begin{figure}
\caption{
Sketch of mode division. $\eta$ and $\eta'$ are frictions.
The arrow with the word 'energy' implies the energy flow.
The effect of $\eta'$ is ignored in the present study.
}
\label{threemodes}
\end{figure}

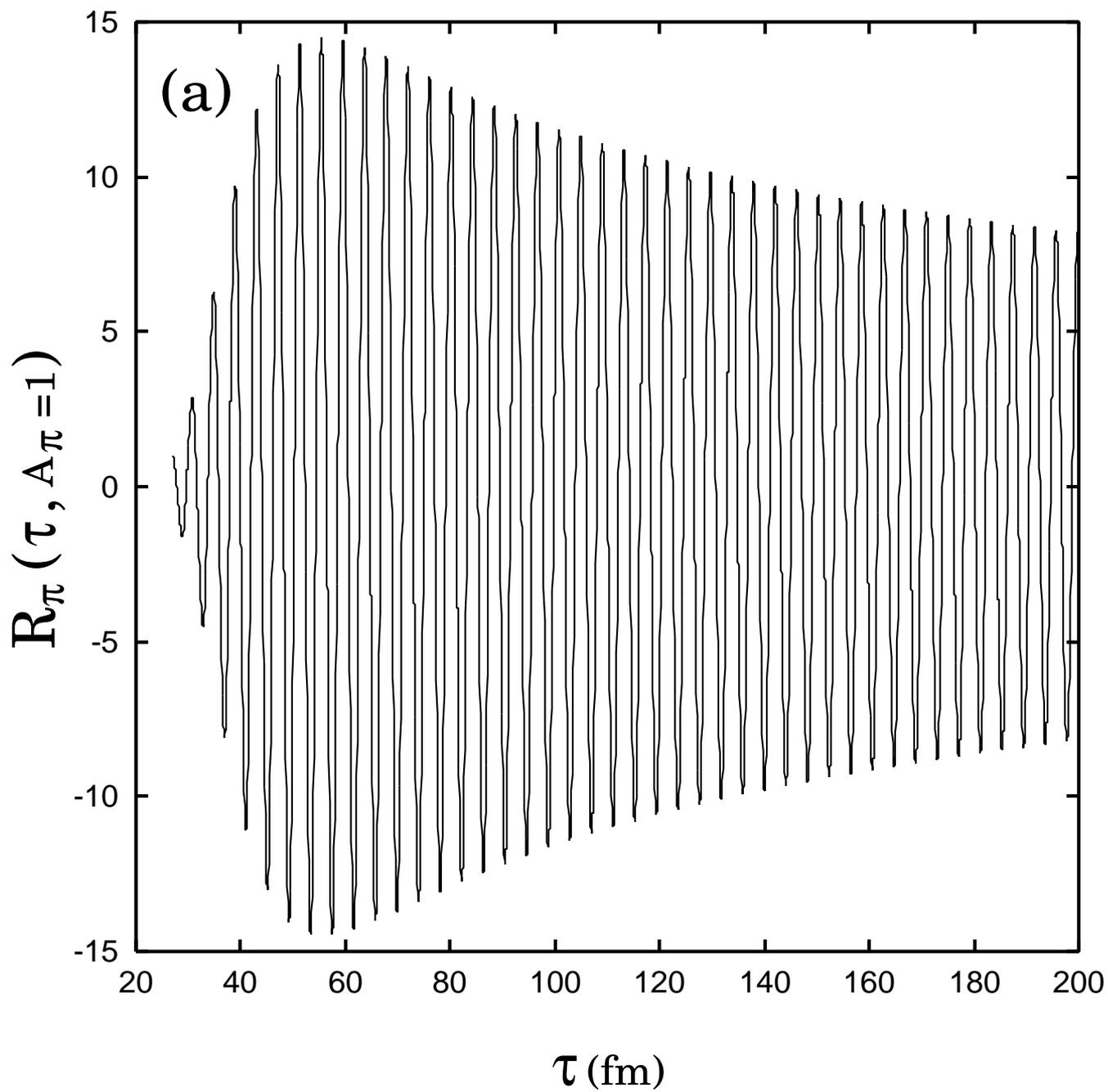
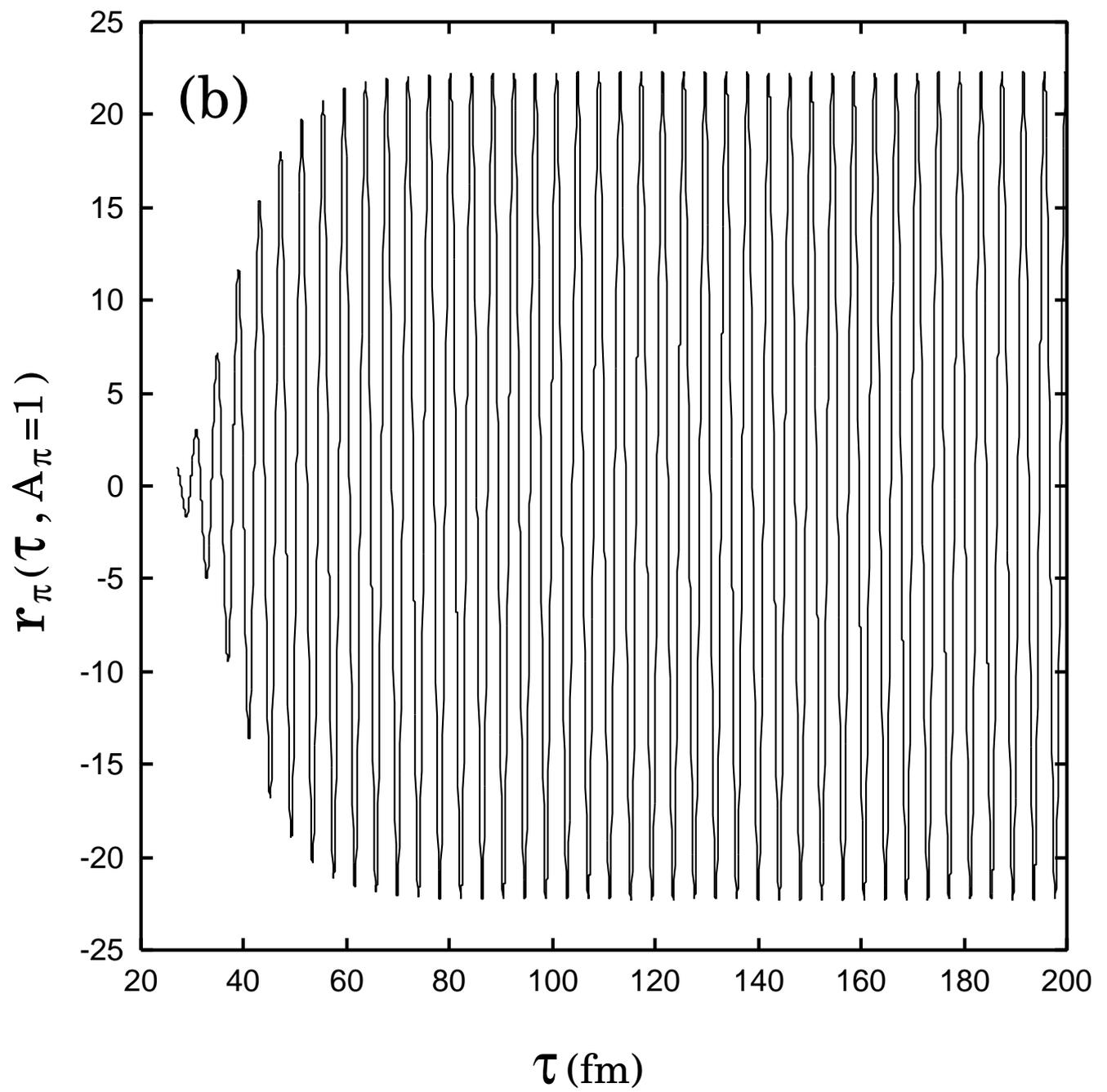
\end{document}